\documentclass[12pt]{iopart}

\usepackage{graphicx}
\usepackage{dcolumn}
\usepackage{multirow}
\usepackage{bm}
\usepackage{amstext}

\def \ee {{\rm e }}
\def \acos {{\rm acos}}

\begin{document}
\title{Irregular collective dynamics in a Kuramoto-Daido system}
\author{Pau Clusella}
\address{Departament de F\'{\i}sica, Universitat Polit\`ecnica de
Catalunya, Campus Nord B4, 08034 Barcelona, Spain}

\author{Antonio Politi}
\address{Institute for Complex Systems and Mathematical Biology, SUPA,
University of Aberdeen, Aberdeen, UK}

\begin{abstract}
We analyse the collective behavior of a mean-field model of phase-oscillators of Kuramoto-Daido type 
coupled through pairwise interactions which depend on phase differences: 
the coupling function is composed of three harmonics.
We provide convincing evidence of a transient but long-lasting chaotic collective chaos, 
which persists in the thermodynamic limit.
The regime is analysed with the help of clever direct numerical simulations, by determining the maximum Lyapunov 
exponent and assessing the transversal stability to the self-consistent mean field. The structure of the invariant measure is 
finally described in terms of a resolution-dependent entropy.

\end{abstract}

\maketitle

\section{Introduction}

The collective dynamics of systems of coupled oscillators has been extensively
studied in the last decades.
In fact, the analysis of macroscopic behavior in complex oscillatory systems appears
in many contexts, neural dynamics being an outstanding example thereof~\cite{Pazo2016,Luccioli-Politi-10,PhysRevX.5.021028,vanVreeswijk1996}.
In order to understand which types of collective regimes are expected in such models,
there is a huge variety of factors that may be important to consider.
Some of them are the complexity of the single-unit dynamics, the shape of the interaction, the topology of the connections, the heterogeneity among the units, and the presence of delayed interactions.
In this labyrinth of possibilities, it is crucial to be aware of which regimes can
be sustained in simple setups, so that they can be used as benchmarks to evetually build a general theory.

In the weak-coupling limit, an ensemble of coupled oscillators can be reduced to a phase model 
where each unit is described by only one variable and
the interaction among them is proportional to the phase response curve, as well as to the 
forcing function~\cite{Winfree1967,Winfree-80}.
A further simplification based on a separation of time scales allows simplifying the structure of the coupling to a function of the phase differences, the so-called Kuramoto-Daido systems~\cite{Kuramoto-84,Daido1993a}.
This family of models provides a framework apt to characterize the emergent dynamics, 
since the macroscopic behavior can be also described in terms of the probability density of the oscillator phases.

Most of the analysis of phase oscillators is centered around sinusoidal coupling: 
this is indeed the form one expects close to bifurcations and has the additional advantage of allowing 
for analytical treatments~\cite{Sakaguchi-Kuramoto-86}. However, over the years it has become increasingly 
clear that sinusoidal coupling is very special: it induces an exactly low-dimensional collective dynamics, 
a property that is immediately lost as soon as 
additional harmonics are added~\cite{Ott-Antonsen-2008,Watanabe-Strogatz-94,Watanabe-Strogatz-93}.
As a result, qualitative changes occur such as the appearance of clusters (otherwise forbidden) 
and the emergence of self-consistent partial synchrony~\cite{Hansel-Mato-Meunier-93,Clusella-Politi-Rosenblum2016}.
Nevertheless, even in the simple case of globally coupled systems, a comprehensive understanding of the
expected dynamical regimes is still missing.
In this paper we address this issue by studing novel dynamical regimes in a Kuramoto-Daido system.

The simplest and most studied regimes arising in ensembles of phase oscillators
are full synchrony and splay states.
They are opposed in the sense that in the former case all oscillators evolve 
with the same phase, whereas the latter corresponds to a uniform distribution of the oscillator phases.
Nevertheless, within both regimes all units display the same dynamics, i.e., there is no symmetry breaking.
Cluster states represent a first instance of symmetry breaking at the microscopic level, but much more interesting are chimera 
states where a symmetry breaking induces the splitting into a synchronous and an asynchronous 
group~\cite{Abrams2004,Kuramoto2002}.
However, the appearance of chimeras requires the introduction of a distance dependent 
coupling\footnote{Or other ``complications", such as additional degrees of freedom in the oscillator dynamics}. 
Chimeras have an additional peculiarity that is important in connection to the regime
we are going to discuss in this paper: they are typically transient states, although 
the transient can be arbitrarily long if the system-size is let diverge~\cite{Wolfrum2011}.
Another family of regimes that has been explored and turns out to be generic is self-consistent-partial synchrony (SCPS)~\cite{vanVreeswijk1996,Clusella-Politi-Rosenblum2016}.
In its simplest form, all oscillators evolve identically, 
but display quasiperiodic motion, 
a more complex evolution than that of the macroscopic observables, which evolve periodically.

The overall scenario is already very broad, but there are still unsolved fundamental questions.
When considering a finite and possibly a small number of oscillators, 
a chaotic dynamics is expected to arisem since, the dynamical system is nonlinear.
This is indeed true and it has even been understood 
that chaos can be observed in as few as four oscillators coupled through phase differences~\cite{Ashwin2011}.
Nevertheless, starting from this setup and increasing the number of oscillators, two scenarios can arise: (i)
either the chaos dilutes itself becoming increasingly weak and eventually disappearing; (ii) clusters form around the chaotic
oscillators giving rise to a pseudo-macroscopic dynamics, i.e., a dynamics where 
the macroscopic chaos is just the consequence of a {\it trivial} arrangement around a few centers.

Is it possible to obtain collective chaos in ensembles of identical phase oscillators?
In the literature, one finds various examples of mean-field coupled systems characterized 
by collective chaos~\cite{Kaneko1990,nakagawa1993,chabanol}.
The closest instance to the Kuramoto-Daido model is a Stuart-Landau ensemble 
in a regime where the oscillators remain confined to a smooth closed
curve and therefore can be parametrized by their phases~\cite{Clusella2019}.
Nevertheless, the amplitude fluctuations are essential for the chaotic dynamics 
and thus this setup does not belong to the class of phase oscillators.
Other instances of collective chaos in phase oscillators require either
heterogeneity or multiple populations~\cite{Olmi2010,Luccioli2010,Bick2018}.

In this paper we discuss a model where collective (macroscopic) chaos persists in the thermodynamic 
limit although it is a transient regime (possibly similar to the transient character of chimera states).
The model we propose makes use of a multimodal coupling function composed of three harmonics.
We have found a set of parameter values where the evolution is characterised by a weak form of collective chaos that 
is robust upon increasing the system size.
The many simulations show that, in fact, we are before a very long 
transient regime, but the collapse is too sporadic for it to be quantitatively explored.

Altogether, in Sec.~2 we introduce the model, while Sec.~3 is devoted to the various methods used to study the system, which include a
clever way to avoid spurious synchronization. In Sec.~4 we present a phenomenological description of different regimes observed upon
varying some control parameter. In Sec.~5 we focus on a more quantitative description, discussing both Lyapunov exponents and the
structure of the invariant measure. In Sec.~6, open problems are discussed.

\section{The model}

We consider an ensemble of $N$ identical, globally coupled Kuramoto-Daido oscillators.
The evolution of each phase $\phi_j\in[0,2\pi)$ is ruled by the equation
\begin{equation}\label{eq:system}
\dot\phi_j=\omega+\frac{1}{N}\sum_{m=1}^N G(\phi_m-\phi_j) \qquad j=1,\dots,N\;.
\end{equation}
Since the system is homogeneous, i.e. all oscillators have the same natural frequency $\omega$,
we are free to choose a rotating frame such that $\omega=0$.
In this paper, we consider a triharmonic coupling function $G(\phi)$
\begin{equation}
        G(\phi):=a_1\sin(\phi+\gamma_1)+a_2\sin(2\phi+\gamma_2)+a_3\sin(3\phi+\gamma_3) \; ,
\end{equation}
where $\gamma_k \in [0,2\pi)$ are the Sakaguchi phase shifts and $a_k>0$ are the harmonic amplitudes.
Through time rescaling, we can assume $a_1=1$, so that only $a_2$, $a_3$, and the phase shifts are
significant parameters for the dynamics of the system.
We choose to work with $a_2=0.8$, $a_3=0.5$, $\gamma_1=1.4$ and $\gamma_3=\pi$, letting $\gamma_2$ to be the
only control parameter to play with.  Such a choice of parameter values will be justified later on.

The dynamics of the phase oscillators can be characterized with the help of the \emph{Kuramoto-Daido order parameters},
\begin{equation}\label{eq2:order-parameters}
 R_k \ee^{i\alpha_k} :=\frac{1}{N} \sum_{j=1}^N \ee^{ik\phi_j} \;,
\end{equation}
where $R_k$ and $\alpha_k$ are real numbers.
In the thermodynamic limit, such order-parameters coincide with the (spatial) Fourier modes
of the corresponding probability density of the phases $P(\phi)$. Hence, they are proper tools to characterize
the macroscopic evolution of the ensemble.
For instance, if the phases are spread uniformly along the unit circle, then $P(\phi)=1/(2\pi)$
and $R_k=0$ for all $k\neq0$. Such state is known as \emph{splay state}.
On the other hand, if all oscillators take exactly the same phase $\phi_j=\psi$ $\forall i$,
then $P(\phi)=\delta(\phi-\psi)$, thus $R_k=1$ $\forall k$. In this case the system is \emph{fully synchronized}.

Using the first three Kuramoto-Daido order parameters, the evolution equation (\ref{eq:system}) can be written as
\begin{equation}\label{eq2:biharmonic-compact}
        \dot\phi_j=  \sum_{k=1}^3 a_kR_k\sin(\alpha_k-k\phi_j+\gamma_k)
\end{equation}
where the mean-field structure of the system becomes evident.

Using these notations, the stability of the simplest regimes can be studied analytically.
In particular, the splay state is stable only in the cube delimited by $\gamma_k\in (\pi/2,3\pi/2)$.
On the other hand, the fully synchronous state is stable in the region delimited by $-G'(0)<0$.
In particular, setting $\gamma_1=1.4$ and $\gamma_3=\pi$ ensures that the splay state
is never stable and that full synchrony is stable only if
\begin{eqnarray*}
        \gamma_2<&\acos\left( \frac{3a_3\cos(\gamma_3)-\cos(\gamma_1)}{2a_2} \right)\simeq 0.589\quad\text{or}\\
        \gamma_2>&2\pi-\acos\left( \frac{3a_3\cos(\gamma_3)-\cos(\gamma_1)}{2a_2} \right)\simeq 5.694
\end{eqnarray*}

In a previous work~\cite{Clusella-Politi-Rosenblum2016}
we showed that in a similar model with a biharmonic coupling function, a non-trivial dynamics emerges
in the parameter regions where neither the splay state nor the fully synchronous states are stable.
Here a similar scenario emerges, but with novel and more complex regimes.

\section{Methods}
\subsection{Numerical integration}
\label{methods:integration}

Numerical integration of system~(\ref{eq:system}) seems a simple task, 
but we must warn the reader that the homogeneity of the system poses important difficulties.
Problems arise when groups of oscillators form a quasi-cluster, with several phases concentrated in a 
very narrow range.
In fact, if the quasi-cluster width becomes smaller than the floating-point accuracy used in the computation, 
then the phases of those units become undistinguishible, thus leading to the permanent formation of spurious cluster states.
Although this problem might seem pathologic, it happens frequently in systems of identical phase oscillators. 
The simplest instance of such artifact arises in the biharmonic model, where
simulations of heteroclinic cycles converge towards a spurious two-cluster 
regime after a long integration time~\cite{Hansel-Mato-Meunier-93,Clusella-Politi-Rosenblum2016}. 
This issue also arises for other types of irregular dynamical behavior 
and it has a major impact in the regimes studied in this paper.

In the literature, two simple techniques are used to overcome such a problem, both of them with some drawbacks. 
One can add a very small amount of heterogeneity in the system, e.g., assigning different natural frequencies 
to the various oscillators.
Another option consists in regularizing the dynamics by adding a small amount of independent 
noise to each oscillator. 
These two methods avoid the formation of spurious clusters, but induce a violation of 
a crucial property of identical determinstic phase oscilators, allowing them to overtake one another.
Additionally, it is known that qualitative changes may arise already for small 
noise~\cite{Clusella2017,Takeuchi2013,Shibata-Chawanya-Kaneko-1999}.

Here, we prefer to follow a different approach, which exploits the invariance of the equations on a
global phase shift.
More precisely, once the oscillator phases have been be ordered so that $\phi_j<\phi_{j+1}$ for $j=1,\dots,N-1$,
we introduce new variables: the distances between consecutive oscillators 
$\delta_j=\phi_{j}-\phi_{j-1}$ for $j=2,\dots,N$ and $\delta_{1}=\phi_{1}-\phi_{N}$.
Making use of Eq.~(\ref{eq2:biharmonic-compact}), it is readily seen that $\delta_j$ obeys the equation 
\begin{eqnarray*}
	\dot\delta_j=  \sum_{k=1}^3 a_kR_k\Bigl[\sin(\alpha_k-&k\phi_j+\gamma_k)(1-\cos(k\delta_j)) \\&- \cos(\alpha_k-k\phi_j+\gamma_k)\sin(k\delta_j)\Bigr]
\end{eqnarray*}
where the Kuramoto order parameters can be computed by recursively determining the actual phases,
$\phi_j=\phi_{j-1}+\delta_j$ for $j=2,\dots,N$.
In principle, one should also keep track of at least one reference oscillator to reconstruct the actual position of all.
As we are interested in the properties of the macroscopic density and since the evolution depends only on
the relative positions, we choose a frame attached to the first oscillator, so that $\phi_1=0$ at all times.
Moreover, since $\sum_{i=1}^N \delta_j=2\pi$, one only needs to integrate $N-1$ equations, 
the last phase difference being given by $\delta_N = 2\pi - \phi_{N-1}$.
However, since very small $\delta_j$ values are expected, the accuracy is lost when the phases are reconstructed.
Therefore, we prefer to integrate all differences, determine the total sum and thereby
rescale phases and distances in order to enforce $\sum_{i=1}^N \delta_j=2\pi$.
With this method we are able to follow pairs of oscillators down to a distance of order $\sim 10^{-323}$.
All results presented in this paper have been obtained using this method.

\subsection{Distribution entropy}
\label{methods:entropy}

In most cases a sequence of snapshots suffices to identify whether
a distribution of oscillators is either a multi-cluster state, or a smooth distribution.
Nevertheless, in some cases the distribution might be smooth with very sharp bumps, or even seemingly fractal.
In order to quantify the complexity of the density distribution from direct simulations of finite systems 
we rely on entropy measures.
Given the ensemble of $N$ oscillators, by definition, $\sum_j \delta_j = 2 \pi$; moreover $\delta_j \ge 0$.
Therefore
\begin{equation}
p_i(m) = \sum_{j=(i-1)m}^{im} \frac{\delta_j}{2\pi}
\end{equation}
can be interpreted as a probability. If we are careful enough to choose $m$ such that $N=mq$, where $q$ is an integer, it
is easy to show that
\[
\sum_{i=1}^q p_i(m) = 1
\]
irrespective of $m$. Given a probability distribution,  we can then determine the corresponding entropy
\begin{equation}
H(m) = -\sum_{i=1}^q p_i \ln p_i\; .
\end{equation}
Then,
\begin{equation*}
\rho\equiv 1/q=m/N
\end{equation*}
can be interpreted as the observational scale.

In particular, if $m=1$ 
\begin{equation}\label{eq:entropy}
	h:=H(1)=-\sum_{i=j}^N \frac{\delta_j}{2\pi}\log\frac{\delta_j}{2\pi}\;.
\end{equation}
In this case, if the probability density of the phases is uniform then  $h = \log N$. On the other hand, if we have 
full synchrony, then $h=\log(1)=0$.
Moreover, for a homogeneous $k$-cluster state, we would have $h= \log k$. Thus $ \exp(h) $ 
quantifies the amount of effectively non-zero distances between oscillators and can be used to discern between
highly clustered states and smooth distributions.

\subsection{Lyapunov exponents}

Lyapunov exponents are a standard tool to analyze chaotic dynamics \cite{Pikovsky-Politi-2016}.
They measure the growth rate of perturbations along a generic trajectory.
In the present context one can distinguish at least between the standard Lyapunov spectrum and the
\emph{transverse} Lyapunov exponents.

\subsubsection{Lyapunov spectrum}

The Lyapunov spectrum $\{ \lambda_j \}$ for $1\le j\le N$, is obtained by determining the growth rate of different volumes along
a given trajectory in the $N$-dimensional phase space. Given the slow convergence, here we shall focus only on the largest one
to establish whether the underlying dynamics is chaotic.
 
In practice, one needs to study the evolution of an infinitesimal perturbation vector
$\boldsymbol{\delta \phi}:=(\delta \phi_1,\dots,\delta \phi_j)$.
Linearization of Eq.~(\ref{eq2:biharmonic-compact}) leads to
\begin{eqnarray*}
	\dot{\delta \phi}_j= \sum_{k=1}^3 ka_k\left[\rho_k \cos(\beta_k-k\phi_j+\gamma_k)
	- R_k\cos(\alpha_k-k\phi_j+\gamma_k)\delta \phi_j\right]\\
\end{eqnarray*}
where
\begin{equation*}
	\rho_k \ee^{i\beta_k}:=\frac{1}{N}\sum_j^N \ee^{ik\phi_j}\delta \phi_j\;.
\end{equation*}
The largest Lyapunov exponent (LE) $\lambda$ is the average growth rate of the norm of $\boldsymbol{\delta \phi}$. 
Practically speaking one can only compute finite-time trajectories: $\Lambda(\tau)$ corresponds to
the finite-time growth rate over a time interval $\tau$ and, accordingly,
\begin{equation*}
	\lambda=\lim_{\tau \to \infty} \Lambda(\tau)\;.
\end{equation*}
Finally, localization properties of the perturbation vector $\boldsymbol{\delta \phi}$,
provide additional relevant information on the system dynamics. We shall do that by referring to the
unit vector
\begin{equation*}
	\boldsymbol{u}=(u_1,\dots,u_N):=\frac{\boldsymbol{\delta \phi}}{\|  \boldsymbol{\delta \phi}\|}\;,
\end{equation*}
where $\| \cdot \|$ denotes the Euclidean norm.

\subsubsection{Transverse Lyapunov exponent}

The transverse Lyapunov exponent $\lambda_T$  measures
the entraintment of a single isolated oscillator forced by the mean-fields generated by a separate coupled system.
In phase oscillators, a single variable is present and thus, there is only one transverse LE which cannot be  positive. 
A negative transverse LE indicates entraintment with the mean-fields, and thus it is expected to characterize
clustered regimes. 
On the other hand, a zero transverse LE indicates disentanglement with respect to the mean-fields and, therefore, 
it is, roughly speaking, expected in regimes without phase-locking such as SCPS or collective chaos. In reality, in
Ref.~\cite{Clusella2019}, we have seen that the scenario can be more complex. We will return to this point 
in Sec.~\ref{sec:charac}.

The linearized equation for the transverse LE reads
\begin{equation*}
	\dot{\delta \phi_j}=-\sum_{k=1}^3 ka_kR_k(t) \cos\left(\alpha_k(t)-k\phi_j+\gamma_k\right)\delta \phi_j
\end{equation*}
where the time-dependence of $R_k$ and $\alpha_k$ is explicitly reported in order to emphasize that they are externally given mean fields. 
In the absense of symmetry breaking, $\lambda_T$ is the same for all $\phi_j$,
thus the index could be dropped, an issue that will be discussed later.
Also in this case one computes the finite-time transverse LE, denoted as $\Lambda_T(\tau)$,
and then use that
\begin{equation*}
	\lambda_T=\lim_{\tau\to\infty} \Lambda_T(\tau)\;.
\end{equation*}

\section{Phenomenology}

In this section we describe the main dynamical regimes observed
in direct simulations of the triharmonic system.

\subsection{Periodic and quasiperiodic macroscopic dynamics}
\begin{figure}[tbp]
  \centerline{\includegraphics[width=0.55\textwidth]{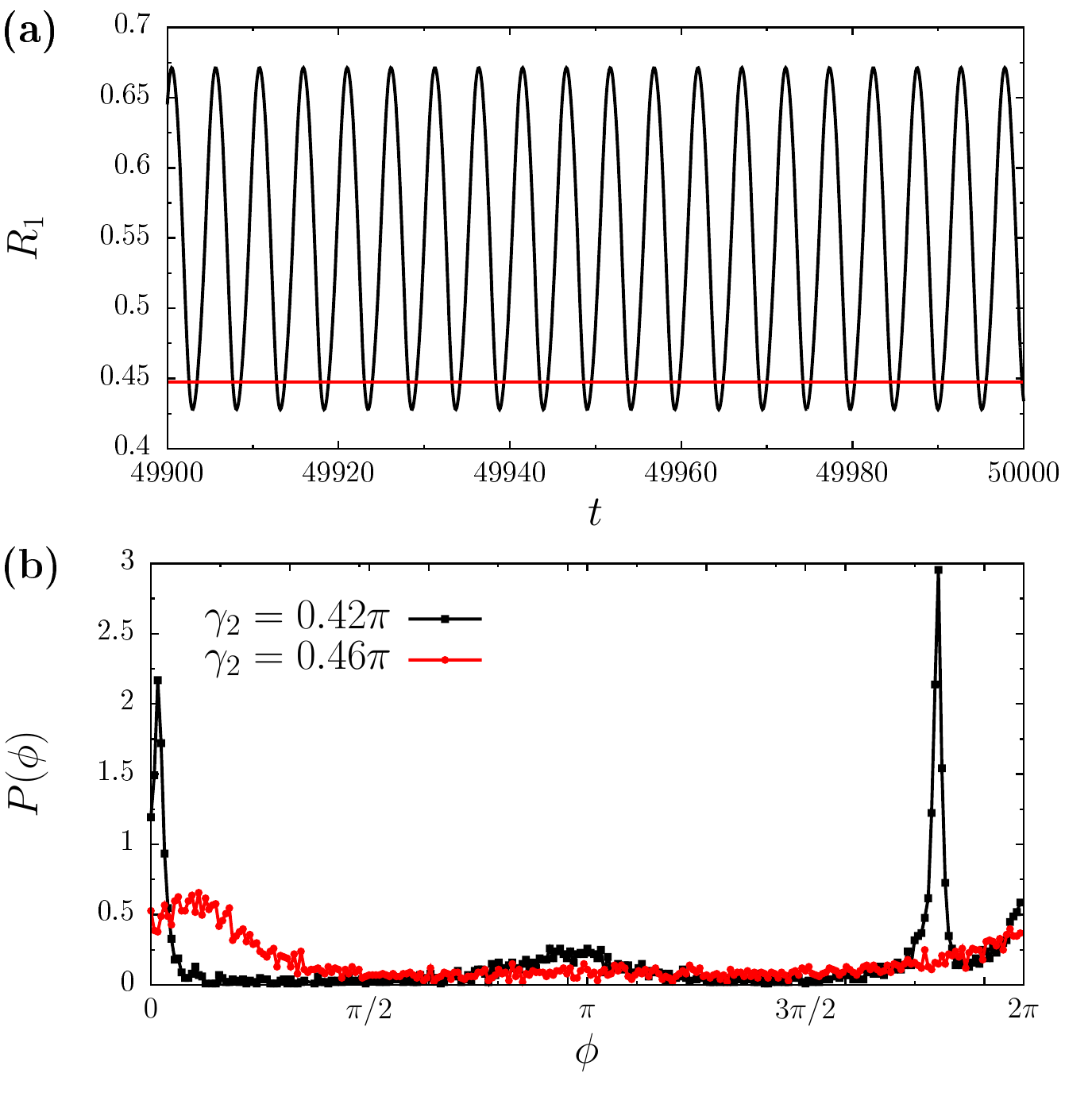}}
  \caption{Results from numerical simulations with $N=4096$ oscillators.
        (a) Time series of the Kuramoto order parameter for $\gamma_2=0.46\pi$ (red) and $\gamma_2=0.42\pi$ (black).
        (b) Normalized histograms of the oscillator positions as obtained from a single snapshot of the simulations,
        red circles correspond to $\gamma_2=0.46\pi$ and black squares to $\gamma_2=0.42\pi$.}
        \label{fig:transition}
\end{figure}

The collective behavior of an ensemble of identical phase oscillators can be characterized
in terms of the probability density of the phases $P(\phi)$, and by
extension, monitoring the Kuramoto order parameters, since, in the thermodynamic limit,
they correspond to the Fourier modes of $P$.
On the one hand, cluster states are characterized by a singular probability density, composed of a collection of $\delta$ functions.
If the cluster distribution is uniform, then $R_k=1$, where $k$ is the number of clusters. In this case there is no
difference between macroscopic and microscopic dynamics, they are both the result of a fixed finite number
of degrees of freedom associated with the position of the clusters.
On the other hand, the splay state is represented by a flat density, $P(\phi)=\frac{1}{2\pi}$, so that $R_k=0$ for all $k>0$.
In this case the macroscopic dynamics is a fixed point, while the single oscillators move periodically and
synchronously.

Following a scale of increasing complexity, periodic self-consistent partial syncrhonization (SCPS) is
a periodic collective regime (the complex Kuramoto order parameter displays a limit-cycle behavior)
accompanied by a quasiperiodic evolution of the single oscillators.
The minimal setup where such state arises is a biharmonic Kuramoto-Daido model studied in \cite{Clusella-Politi-Rosenblum2016}.
The triharmonic model can be seen as an extension of such model and thus it is not surprising to see that SPCS emerges
for similar parameter values of the first two harmonics.
In particular, choosing $\gamma_1=1.4$, $\gamma_2=\pi$, and $\gamma_3=\pi$  neither the splay state nor the full synchronous regime are stable,
and numerical simulations of system (\ref{eq:system}) reveal periodic SCPS\footnote{A further reduction of $\gamma_1$ can trigger the destabilization of SCPS and the appearance
of heteroclinic cycles between two two-cluster states in a similar fashion as it happens in \cite{Clusella-Politi-Rosenblum2016}.}.
This regime persists from $\gamma_2=\pi$ to $\gamma_2\sim 0.459\pi$ for $\gamma_1=1.4$ and $\gamma_3=\pi$.
The red line in Fig.~\ref{fig:transition}(a) shows a time series of $R_1$ for $\gamma_2=0.46\pi$, that is constant and slightly below
0.45, indicating an intermediate level of synchronization.
Complementing this information, an instantaneous histogram of the phases is shown in Fig.~\ref{fig:transition}(b) (see the red curve),
that is rather flat and with a single bump.
As expected for periodic SCPS regimes (finite-size fluctuations left apart), 
the density distribution is rather smooth and rotates rigidly
with a characteristic period, i.e. it is stationary in a suitably rotating frame (the information on the frequency is
contained in the phase of the complex Kuramoto order parameter).

Around $\gamma_2\sim 0.459\pi$ there is a torus bifurcation and a quasiperiodic SCPS sets in.
The complex Kuramoto order parameter now behaves quasiperiodically, while the modulus $R_1$ displays periodic oscillations,
as shown by the black curve in Fig.~\ref{fig:transition}(a).
The distribution of the phases is no longer stationary and displays additional bumps as it can be
observed in the histogram depicted by black squares in Fig.~\ref{fig:transition}(b).
Quasiperiodic SPCS persists until it dies out around $\gamma\simeq 0.4011\pi$.
This type of torus bifurcation from periodic to quasiperiodic SCPS was previously
identified in systems of homogeneous quasi-phase oscillators \cite{nakagawa1995,Clusella2019}
and networks of QIF neurons with synaptic delay \cite{Devalle2018}.
This is the first instance of its appearance in a system of identical phase-oscillators without delay.

\subsection{Irregular dynamics}
\begin{figure}[tbp]
  \centerline{\includegraphics[width=0.5\textwidth]{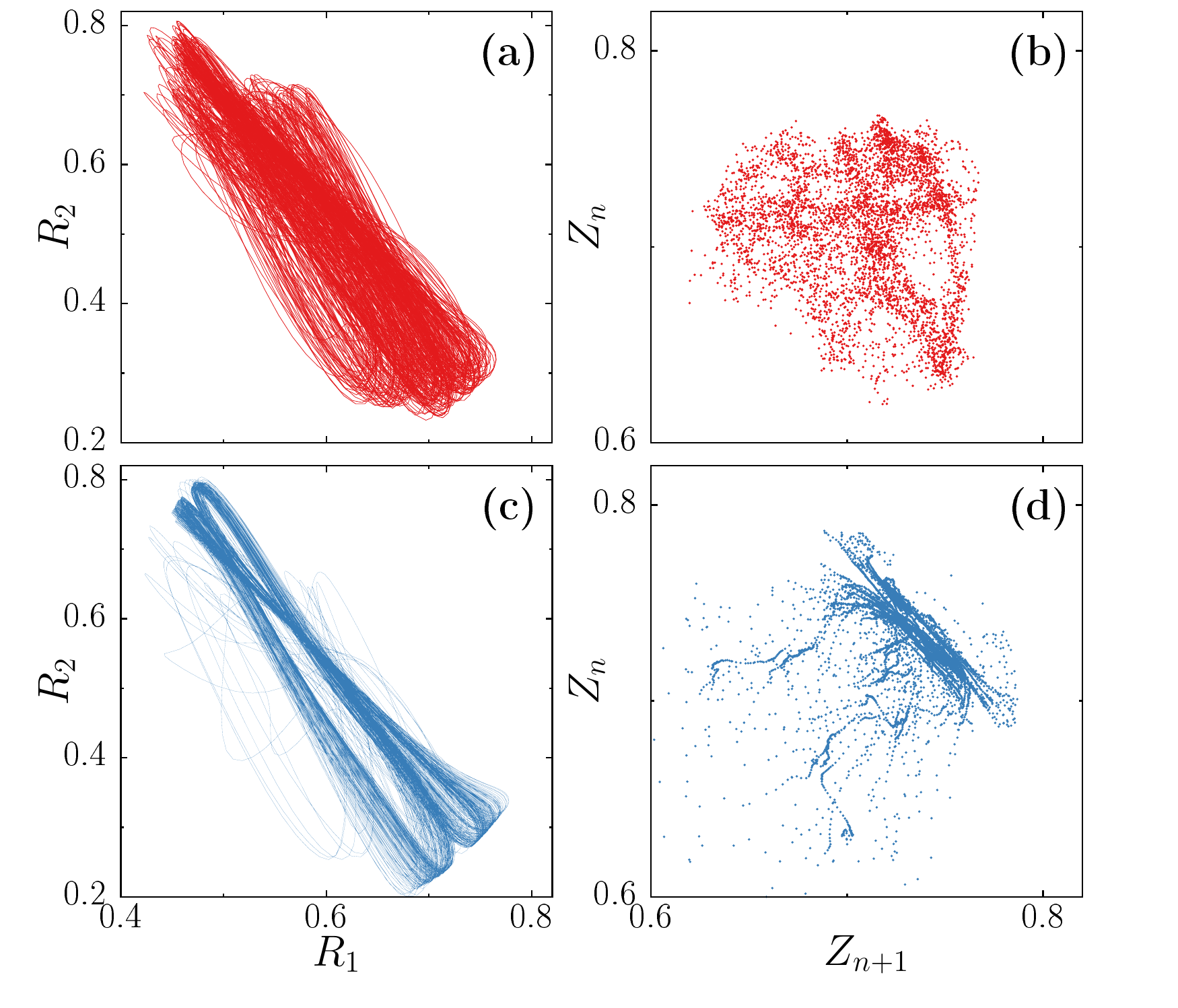}}
  \caption{Results from numerical simulations with $N=4096$ oscillators. Panels (a) and (b) correspond to the chaotic regime, whereas panels (c) and (d) correspond to the intermitent clusters.
(a,c) Coevolution of the first two Kuramoto order parameters $R_1$ and $R_2$ showing the irregularity of the dynamics for both cases.
(b,d) Poincar\'e section of the first Kuramoto order parameter.}
        \label{figTS}
\end{figure}

Above $\gamma_2\sim0.4011\pi$ the macroscopic evolution of the system becomes intricate.
From now on, we set $\gamma_2=0.4\pi$ as an instance of a typical regime
and study the behavior of the system in detail.
For these parameter values, a perfect anti-phase cluster regime is stable and coexists with two other
regimes we are going to focus on. 
One way to avoid a collapse on the two-cluster state is by choosing a unimodal initial distribution of phases 
such as a wrapped Lorenzian or a wrapped normal distribution.

We illustrate the two regimes by commenting the outcome of numerical simulations performed with $N=4096$.
In both cases the evolution of the Kuramoto order parameters is erratic and snapshots of the phase distributions
are highly clustered in a few groups (from 7 to 9, depending on the simulation).
Figure~\ref{figTS}(a) and (c) show the projection of the trajectories in the plane spanned by $R_1$, $R_2$
for the two regimes that we call A and B, respectively.
The two projections are somehow similar and reveal the presence of an irregular dynamics.
The difference between the two regimes is more evident in Fig.~\ref{figTS}(b) and (d), where we plot a
Poincar\'e sections of the Kuramoto order parameter, built as
\begin{equation*}
	Z_n=R_1(t_n)
\end{equation*}
where $t_n$ are the time points where $R_1$ reaches a local maximum.
The section corresponding to regime A (Fig.~\ref{figTS}(b)) is a blurred cloud of points with little structure,
suggestive of a chaotic evolution of the system.
On the other hand, the section corresponding to regime B (Fig.~\ref{figTS}(d)) shows irregular lines, corresponding to
oscillatory trajectories of the Kuramoto order parameter with a smooth variation of the amplitude.
Thus, we are still in front of a complex dynamical regime, but more regular than A.

\begin{figure*}[tbp]
  \centerline{\includegraphics[width=1\textwidth]{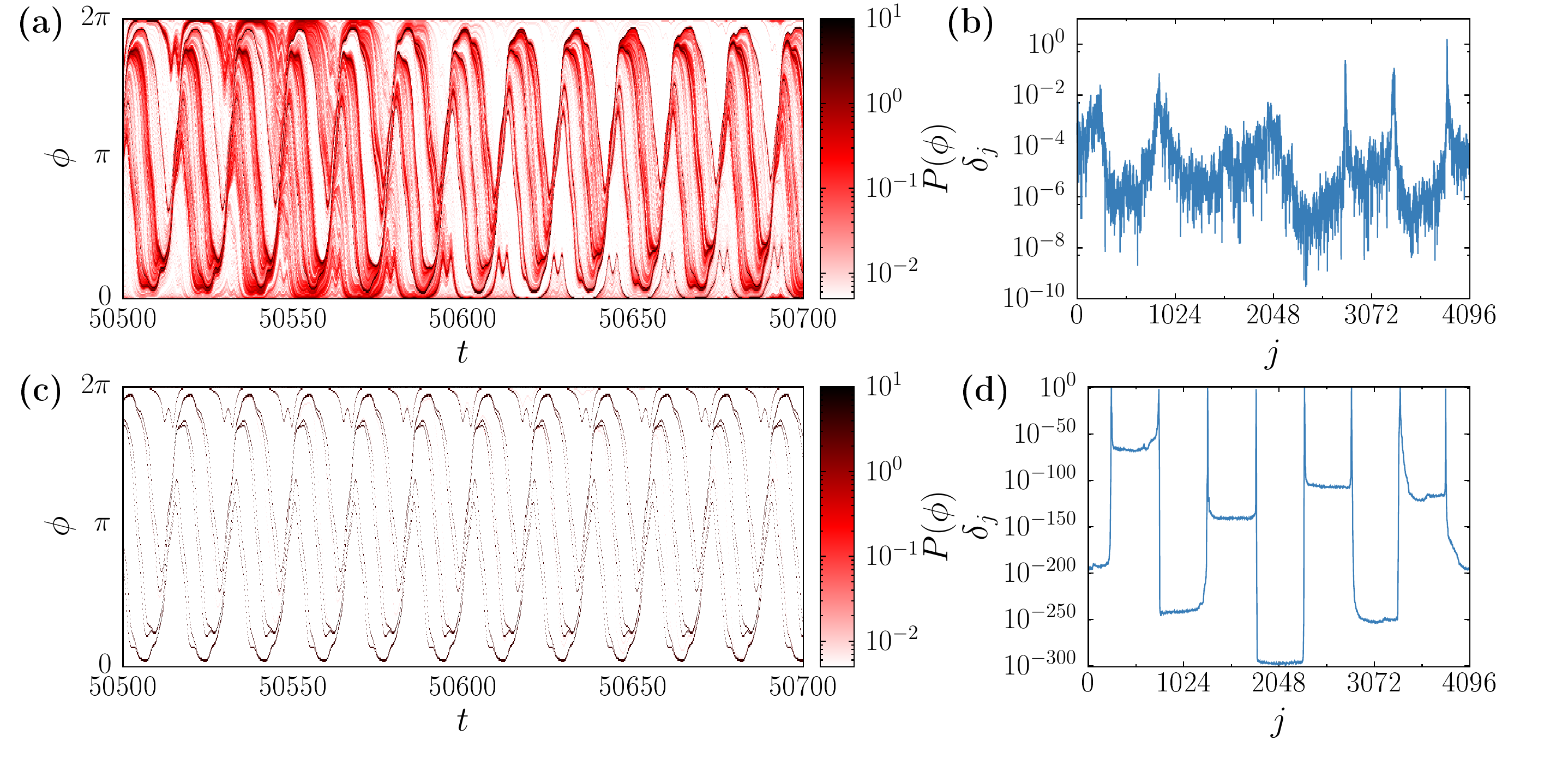}}
  \caption{Results from numerical simulations of a system with $N=4096$ units.
Panels (a) and (b) correspond to the chaotic regime, whereas panels (c) and (d)
correspond to the intermitent clusters.
(a,c) Time evolution of the distribution of the phases obtained as a normalized histogram.
(b,d) Snapshot of the distances between oscillators at $t=5\times 10^4$.}
        \label{figQual}
\end{figure*}

In order to gain further insight on both regimes, we leave aside the macroscopic observables and focus on the evolution
of the single oscillators. In Fig.~\ref{figQual}(a) and (c) we plot the dynamics of the density (the amplitude is color
coded as quantified by the side bar)\footnote{The reference frame is selected according to the procedure described
in the previous section, i.e. it is attached to a randomly chosen oscillator labelled as $\phi_1$.}.
Panel $(a)$ refers to regime $A$; the distribution contains a few sharp bumps or quasi-clusters
(notice the logarithmic scale of the density).
A snapshot of the oscillator distances $\delta_j$ allows for a more quantitative characterization.
In Fig.~\ref{figQual}(b), one can see that the $\delta_j$-values indeed cover a wide range of scales but are nevertheless
larger than $10^{-10}$, suggesting a finite cluster width.
On the other hand, the simulation of regime B, displayed in figure \ref{figQual}(c), reveals the presence of
almost perfect clusters. This impression is confirmed by the snapshot presented in
Fig.~\ref{figQual}(d), where we see plateaux heights (which correspond to the single clusters) 
as small as $10^{-300}$.
The large gap between the plateaux heights is the consequence of a mechanism similar to what observed in
heteroclynic cycles~\cite{Hansel-Mato-Meunier-93,Clusella-Politi-Rosenblum2016}: each cluster alternates long periods of
stability (when mutual distances shrink) with periods of instability (when distances grow exponentially).
Once the width of an exploding quasi-cluster is large enough, the distribution suffers major changes which end up in
a switch between exploding and shrinking clusters.
These breakdowns or switching events are responsible for most of the irregularities
of the system shown by the trajectory of the Kuramoto-order parameter and visible in the Poincar\'e section
(Figs.~\ref{figTS}(b) and (c)).

\begin{figure}[t]
  \centerline{\includegraphics[width=0.6\textwidth]{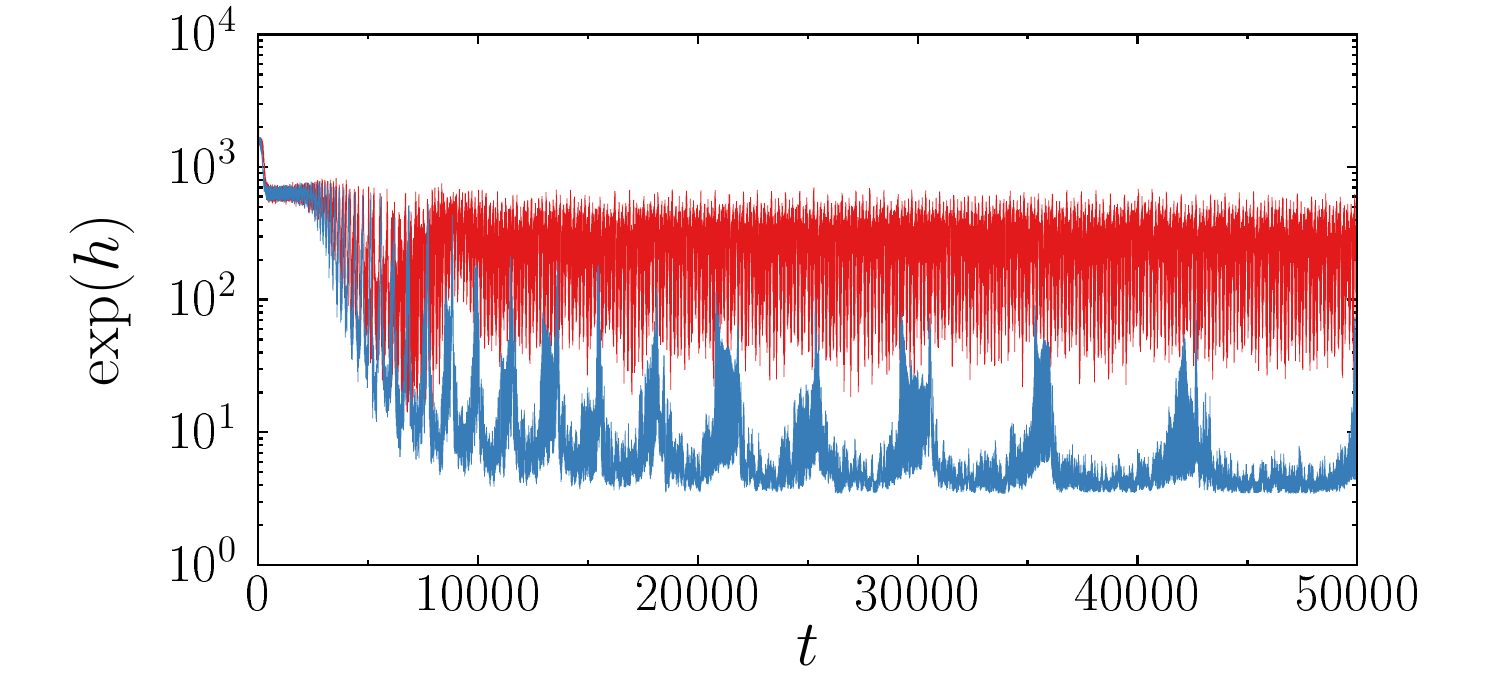}}
	\caption{Time series of the exponential distribution entropy $\exp(h)$ for the chaotic state (red) and the intermitent regime (blue) with $N=4096$.}
        \label{figEntropy}
\end{figure}

A more quantitative analysis of the two regimes can be performed by computing the distribution entropy $h$ as from Eq.~(\ref{eq:entropy}).
Since the shape and the overall structure of the phase density fluctuates, it is convenient to look at its time dependence.
Fig.~\ref{figEntropy} shows a time series of the exponential entropy $\exp(h)$ for both regimes starting from time 0.
It is useful to remind that $\exp(h)$ represents the  number of sites where the density is effectively 
different from zero. In fact, it can range from the minimal value 1, for a single-cluster distribution,
to a maximal value equal to the number $N$ of oscillators ($N=4096$ in our case) for a flat distribution.
In the case of regime A (red curve), after a relatively short transient, a seemingly stationary regime sets in,
characterized by relatively large fluctuations, which cover about a decade of  values (from $~50$ to $~600$).
On the other hand, in regime B, $\exp(h)$ takes much smaller values (blue curve), becoming smaller than 10, a 
value which corresponds to the actual number of clusters. Sporadically, bursts are observed, where $\exp(h)$ can become ten times larger:
they correspond to the above mentioned cluster breakdowns.
Such events become increasingly rare  with time, since the width of the quasi-clusters keeps reducing.
In fact, although these simulations have been computed using the method described in section \ref{methods:integration},
regime B ends up in a spurious cluster-state, since our method, though very accurate, is not able to handle distances smaller than $10^{-323}$.

Summing up, we are before two different irregular dynamical types of collective dynamics.
Regime B consists of  a collection of quasi-clusters that are not transversally stable,
thus displaying breathing phenomena.
The main mechanism of switching stability between clusters is akin to that of the heteroclycnic cycles, although
the scenario is more complicated because of the larger number of quasi-clusters. Regime B seems
to be an instance of chaotic itinerancy~\cite{Kaneko90b}.
A more precise characterization is, however, necessary to support this conjecture.
From now on, we prefer to focus on regime A,
as it is a good candidate for representing the first instance of a collective chaotic dynamics in an ensemble of identical phase oscillators.

\begin{figure}[t]
  \centerline{\includegraphics[width=0.6\textwidth]{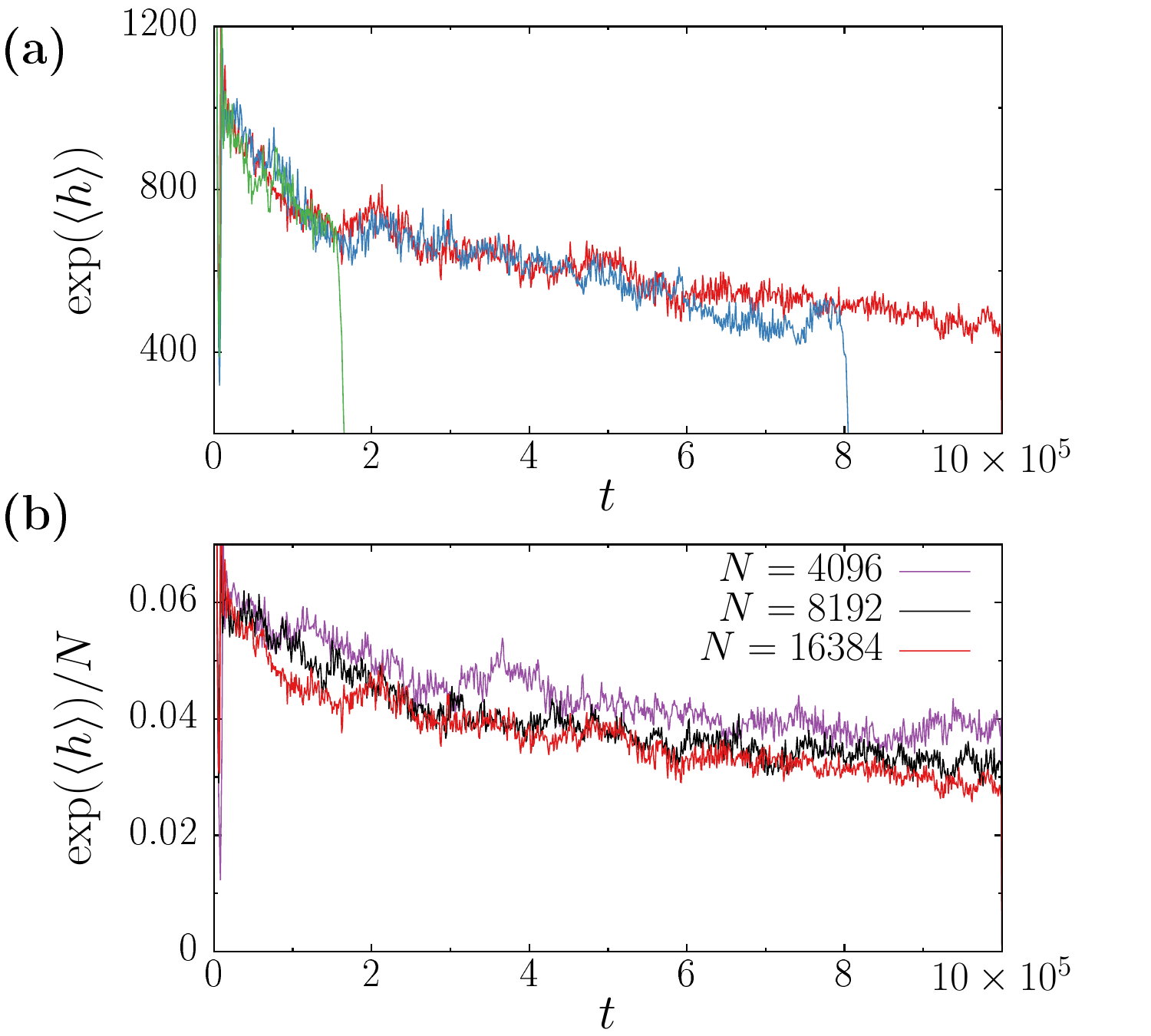}}
\caption{(a) Time series of the exponential entropy, $\exp(\langle h\rangle )$, corresponding to three independent 
simulations with $N=16384$ oscillators each. The exponential entropy $\langle h\rangle$ has been averaged over time windows of $10^3$ time units in order to capture the trend.
(b) Rescaled exponential entropy for three
different ensemble sizes: $N=4096$ (purple), 8192 (black), and 16384 (red).}
\label{figTransient}
\end{figure}

We conclude this section with a brief analysis of the robustness of regime A.
In order to capture only the long-term trend  of $h$ and reduce its time fluctuations
we compute $\exp(\langle h\rangle)$ where $\langle h\rangle$ is the entropy averaged over time windows of $t=10^3$ time units.
Figure~\ref{figTransient}(a) shows long time series of the exponential entropy $\exp(\langle h\rangle)$ for three
independent simulations with $N=16384$, initially falling onto regime A.
In one case, the irregular regime persists more than $10^6$ time units (see red line), but in the two other cases
a sudden crisis leads to a collapse onto regime B, as revealed by the
sharp decrease of the entropy (see green and blue curves in Fig.~\ref{figTransient}(a)).
Given the length of the simulations, it is too much time consuming to determine the average transient time and
its dependence on the system size. We rather focus on the characterization of the transient dynamics itself,
which is already a hard task. In fact, we see that the regime prior the data collapse is not entirely stationary
as the exponential entropy shows a slow decrease with time.
In Figure~\ref{figTransient}(b) we plot the rescaled $\exp(\langle h\rangle)/N$ for three different system sizes ($N=4096$, 8192, and 16384).
Apart from the unavoidable statistical fluctuations, the three curves reveal a substantial independence of $N$,
which suggests that finite-size corrections are negligible.
More important is the time variation: it is plausible to conjecture that 
$\exp(\langle h\rangle)/N$ remains finite for $t \to \infty$, but we cannot 
exclude that the asymptotic regime is characterized by fractal distribution
(we return to this point in the next section).  

\begin{figure}[t]
  \centerline{\includegraphics[width=0.45\textwidth]{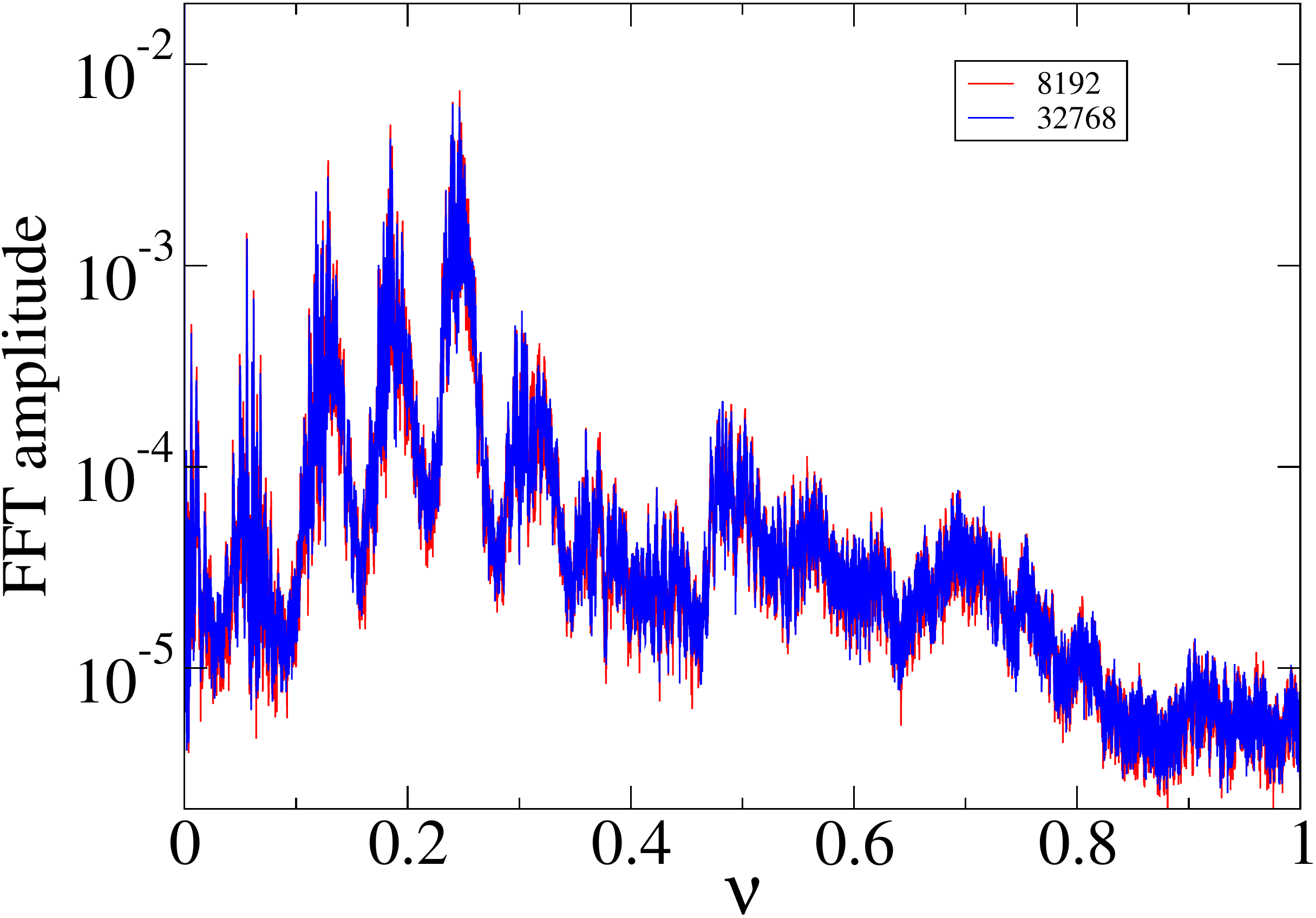}}
  \caption{Amplitude of the Fourier spectrum for two different network sizes. Simulations are performed over $10^5$ time units
(an average over 10 consecutive channels is performed to smooth the spectra). }
        \label{figPSD}
\end{figure}

Finally, in Fig.~\ref{figPSD} we plot the power spectra obtained 
from the evolution of $R_1(t)$ for two different system sizes (8192 and 32768). 
The two spectra nicely overlap over the whole frequency range, suggesting that the dynamics is not affected by
(appreciable) finite-size corrections.
The spectra are concentrated around a few characteristic frequencies.
It is difficult to distinguish harmonics from fundamental frequencies, but the most important feature is the
size-independent width which hints at a truly chaotic dynamics in the thermodynamic limit.

\section{Characterization of the chaotic dynamics}\label{sec:charac}
In this section we perform a quantitative analysis of the collective dynamics, 
focussing on the quasi-stationary regime A identified in the previous section.
We start by investigating its symmetry properties, then proceed to perform a linear stability and finally analyse
the invariant measure.

The phenomenological analysis presented in the previous section has revealed the presence of quasi-clusters, i.e.  a strong
localization of the probability density. The presence of perfect clusters would mean that the symmetry among all oscillators is
broken: some oscillators belonging to a specific subset would share their 
dynamical properties only with the other elements of the
same cluster. The scenario is different in SCPS, where each oscillator progressively visits low- as well as high-
density regions; more formally, all oscillators explore ergodically the same portion of phase space. 

Here a direct detailed test is not possible, because of complexity of the dynamics. One can, nevertheless, devise a
simple test, which allows spotting the occurrence of a symmetry breaking.
In practice, we monitor the distance $\delta_j$ between consecutive oscillators and compute the rescaled average
\begin{equation}\label{eq:ave_dis}
d_j(t) =  \frac{N}{2\pi} \langle \delta_j \rangle \; ,
\end{equation}
where $\langle \cdot \rangle$ denotes a time average from time 0 to time $t$.
If all oscillators behave in the same way (no symmetry breaking), then $\lim_{t\to\infty} d_j = 1$.
By definition, the ensemble average of $\delta_j$ is equal to $2\pi/N$ at each time step, since
they are distributed along the ring $[0,2\pi)$ and the ordering of the oscillators 
does not change over time. The question is whether time averages coincide with the ensemble average. 

\begin{figure}[t]
  \centerline{\includegraphics[width=0.5\textwidth]{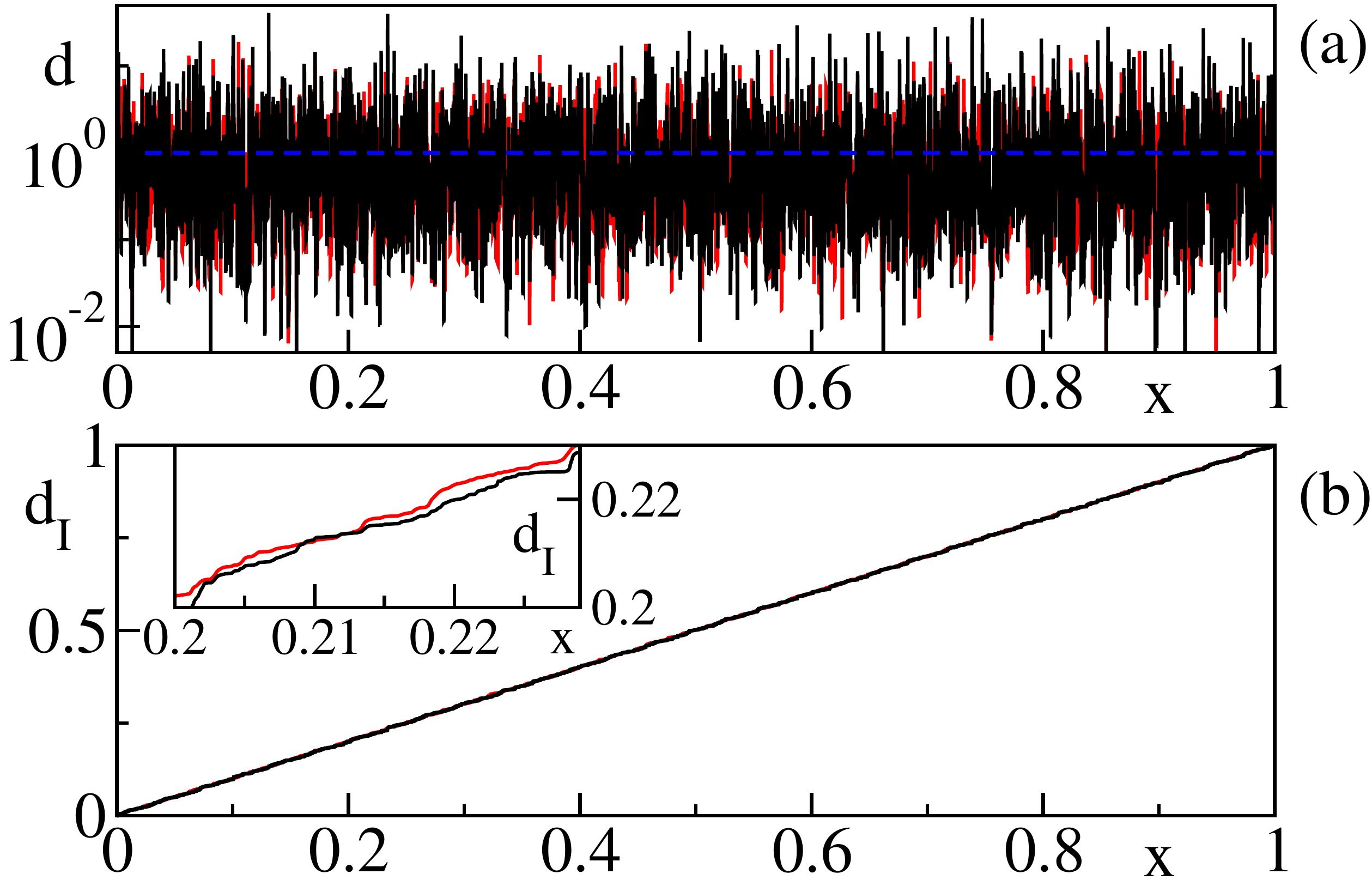}}
  \caption{Simulations performed with $N=8192$ (red) and $N=32768$ (black). Panel (a):
Average distance of neighbouring oscillator rescaled as in Eq.~(\ref{eq:ave_dis}) versus the position $x=i/N$.
The average performed over $10^5$ time units.
In panel (b), the integral $D(x)$ of $d(x)$ is displayed, exhibiting a nearly perfect linear growth.
Fluctuations can only be seen by enlarging the picture as shown in the inset.}
        \label{fig:ave_dis}
\end{figure}

In Fig.~\ref{fig:ave_dis} we plot $d(x) = d_{j/N}$, averaged over a time span $\Delta t = 10^5$ time units,
as a function of the oscillator label rescaled to the system size ($x=j/N$). Red and black curves correspond to
$N=8192$ and $N=32768$, respectively.
The large fluctuations (the vertical scale is logarithmic) around the average value $d=1$ challenge the equivalence
between the different pairs of oscillators.

More useful information emerges from panel (b), where we plot the integral
$d_I(x)$ of $d(x)$ for the same two network sizes. Both curves are basically indistinguishable from straight lines: one
has to look at tiny scales to resolve fluctuations (see the inset). Accordingly, it is natural to conjecture that
symmetry is preserved, but the time scale for the fluctuations to become vanishingly small is exceedingly large.
This hypothesis is consistent with previous stability analyses of asynchronous states and SCPS in chains of phase oscillators
\cite{Clusella-Politi-Rosenblum2016}, which reveal that the convergence rate of high spatial frequencies is exponentially small. In other words, high spatial frequencies are presumably affected by a very weak stability.
We believe that this phenomenon is responsible also for 
the slow convergence of the exponential entropy observed in the previous
section.

\subsection{Lyapunov analysis}
\begin{figure*}[t]
  \centerline{\includegraphics[width=1\textwidth]{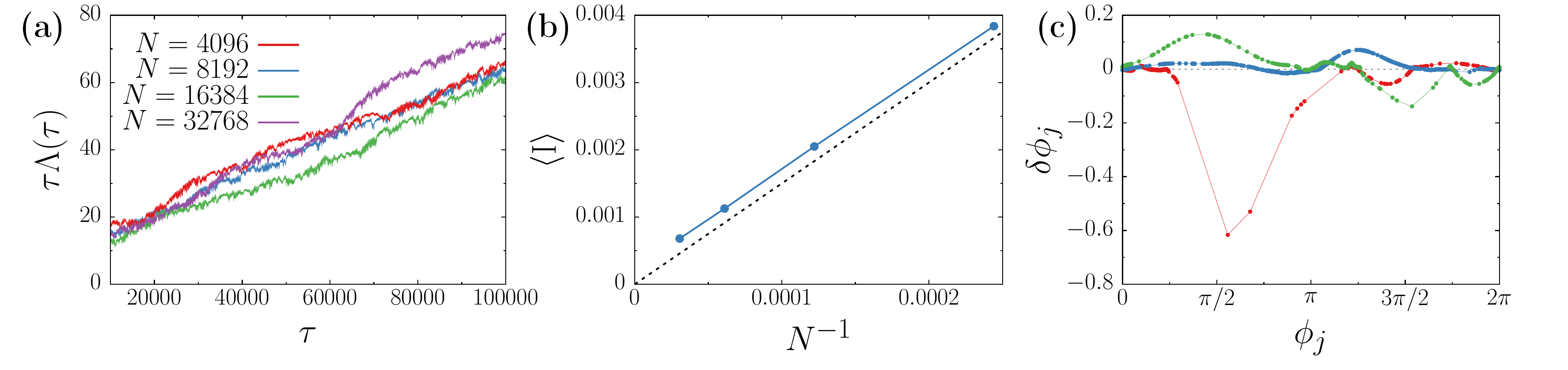}}
  \caption{(a) Integrated growth rate of a perturbation along the most expanding direction for different system sizes.
(b) Time-averaged inverse participation ratio showing its $1/N$ linear dependence.
(c) Instantaneous snapshots of the Lyapunov vector for $N=8192$.}
        \label{fig:lyap}
\end{figure*}

In the previous section we have seen that the collective dynamics is irregular at the macroscopic level. What about the microscopic
dynamics? The most powerful tools for a detailed characterization is linear stability analysis. In Fig.~\ref{fig:lyap} we plot
various indicators associated with the evolution of infinitesimal perturbations.
In Fig.~\ref{fig:lyap}(a), the cumulative expansion rate $\tau\Lambda$ is reported for four different system sizes over a
time span of $10^5$ units. Their slope corresponds to the maximum Lyapunov exponent which turns out to be rather small,
$\lambda \approx 7\cdot 10^{-4}$. The curves for the different $N$ values are relatively close to one another, suggesting that
$\lambda$ is insensitive to finite-size effects.

Additional information can be extracted from the structure of the corresponding Lyapunov vector, in particular from its localization
properties. It has been speculated that the presence of collective chaos (as we indeed observe) might be associated with the
presence of (some) extended Lyapunov vectors~\cite{Takeuchi2013}.
Hence, we have computed the inverse participation ratio $I$. Given the unit norm Lyapunov vector $\boldsymbol{ u}$, the inverse participation ratio is defined as
\begin{equation}
I(N) = \sum_{j=1}^N u_j^4
\label{eq:IPR}
\end{equation}
If $I(N)$ remains finite for $N\to \infty$, one concludes that the vector is localized, while $I \propto 1/N$ means that the vector is
extended. Since the orientation of the vector changes over time, it is convenient to look at the time average $\langle I \rangle$.
The results are presented in Fig.~\ref{fig:lyap}(b), where we see that $\langle I\rangle$ grows linerly with $N$ although the 
proportionality constant
is quite small, suggesting that only 6\% of the sites are characterized by a nonnegligible amplitude.
Direct evidence of extensivity is clearly visibile in panel (c), where we plot a few snapshots of the Lyapunov vectors.
Rather than reporting the amplitude as a function of the oscillator label,
we prefer to use the position (angle) as the independent variable.
This way, the representation has a direct physical interpretation, as the 
amplitude of the perturbation is associated to the position along
the ring. The relative smoothness of the  ``curves" confirm the presence of an extensive structure.
The jumps we see here and there are due to the sporadic occurrence of large gaps in the distribution of phases. We 
conjecture this to be a finite-size effect (see the next subsection).

\begin{figure}[t]
  \centerline{\includegraphics[width=0.45\textwidth]{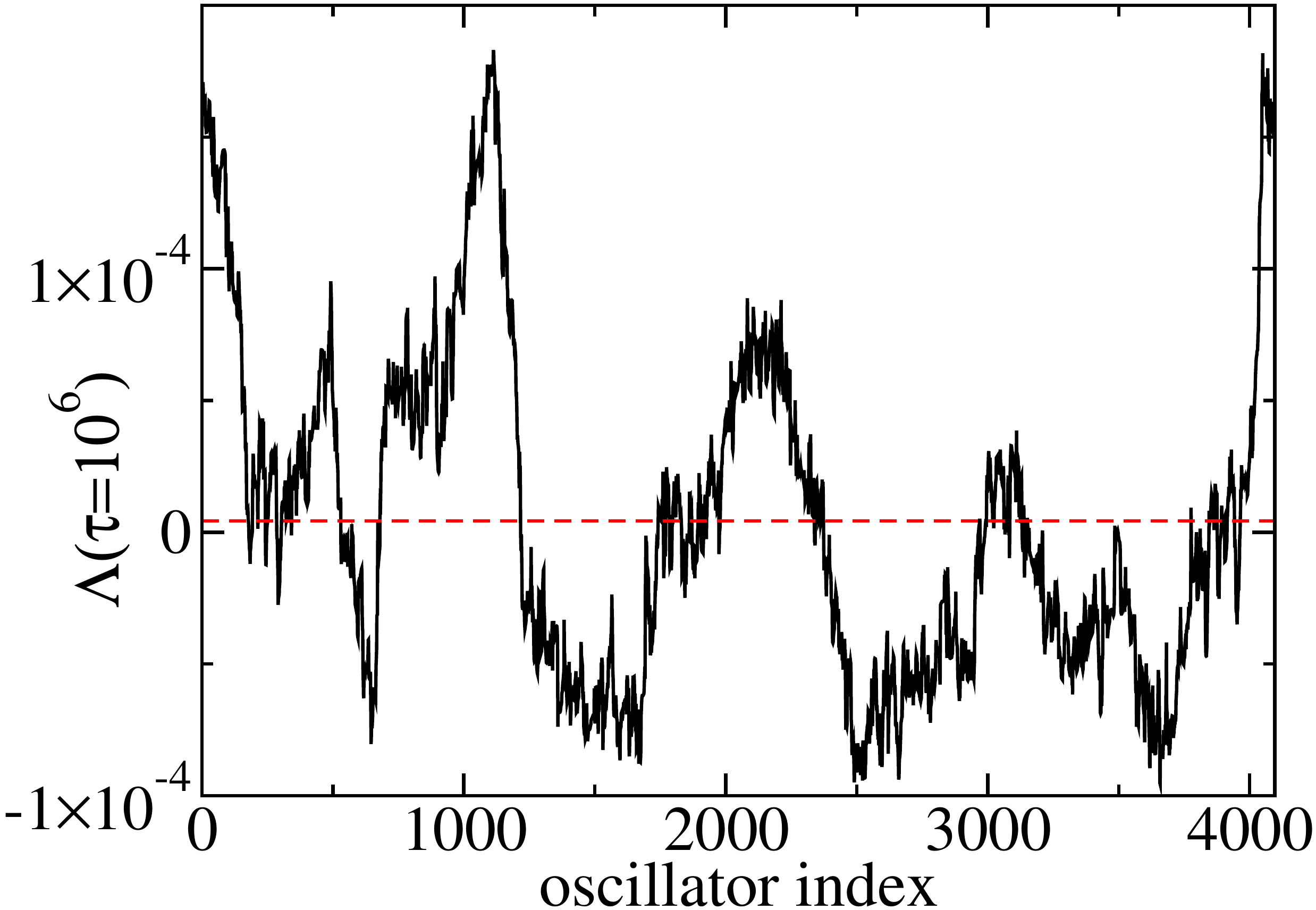}}
  \caption{Finite-time transversal Lyapunov exponent computed over a time span $\tau = 10^6$ versus the oscillator label
in an ensemble of 4096 oscillators. The horizontal dashed line corresponds to the average exponent.}
        \label{fig:lyap_transv}
\end{figure}

The transverse Lyapunov exponent is another useful indicator.
In Fig.~\ref{fig:lyap_transv}, we plot $\Lambda_T(\tau)$ computed over $\tau= 10^6$ time units.
The ensemble average corresponds to
the horizontal dashed line, which is basically indistinguishable from zero.
We are, therefore, led to conclude that the
transverse Lyapunov exponent $\lambda_T = \lim_{\tau=\infty} \Lambda_T(\tau)=0$.

A zero $\lambda_T$ is superficially consistent with the absence of clustering phenomena, which necessarily
imply a negative exponent. However, the relationship between the transverse Lyapunov exponent 
and the singularity of the associated probability distribution is more subtle than one might naively think.
When the collective dynamics is irregular, $\Lambda_T(\tau)$ fluctuates as confirmed by the data plotted in Fig.~\ref{fig:lyap_transv};
this typically implies the existence of an entire range of (transverse) Lyapunov exponents as captured by multifractal formalism,
\[
\mathcal{L}(q) = \lim_{\tau \to \infty} \frac{1}{q\tau}\ln \langle \mathrm{e}^{\Lambda_T(\tau)q}\rangle
\]
where $\langle \cdot \rangle$ denotes an ensemble average over different trajectories.
$\mathcal{L}(0) = \lambda_T$, while $\mathcal{L}(1)$ corresponds to the expansion of linear lengths
(it coincides with the topological entropy in 2d maps).
In \cite{Clusella2019} it was speculated that the general condition for a non-clustered density is not $\mathcal{L}(0)=0$, but $\mathcal{L}(1)=0$.
In the case of a regular collective dynamics such as SCPS, the two conditions coincide since there are no fluctuations and
$\mathcal{L}(q)$ is thereby independent of $q$.
In our case, there are fluctuations and we need to estimate their effect.

\begin{figure}[t]
  \centerline{\includegraphics[width=0.45\textwidth]{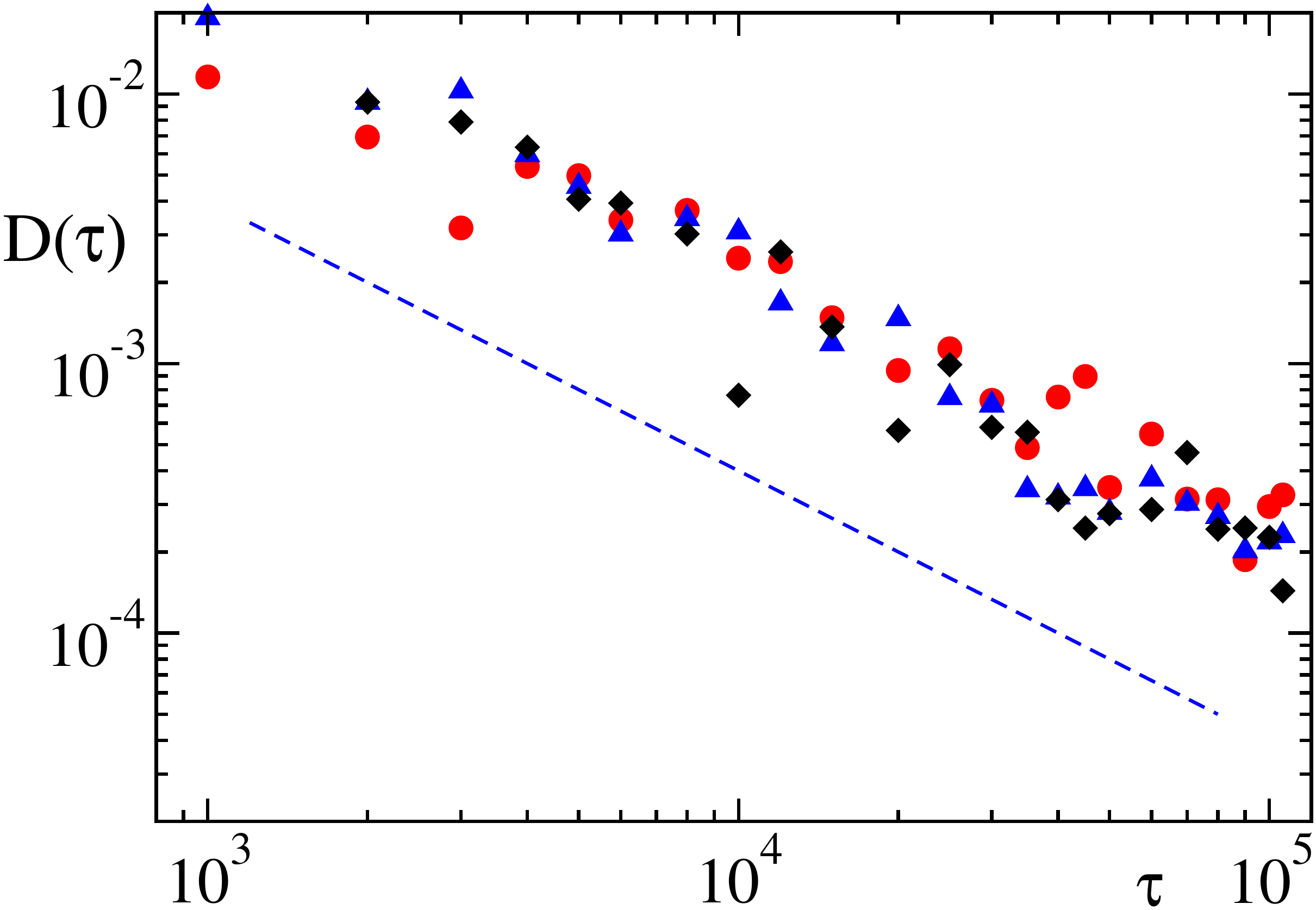}}
  \caption{Fluctuations of the finite-time transverse Lyapunov exponents: the effective diffusion constant $D(\tau)$ is plotted
versus time for three different system sizes (full dots, triangles and diamonds correspond to $N=4096$, 8192, and 16384, respectively). $D$ is determined by averaging over all oscillators. The straight line corresponds to a $1/\tau$ decay.}
        \label{fig:lyap_t}
\end{figure}

Rather than computing $\mathcal{L}(1)$ (hard calculation, because it implies dealing with averaging rare events in 
the limit $\tau \to \infty$), here we rely on the perturbative formula~\cite{Clusella2019}
\begin{equation}
\mathcal{L}(1) = \lambda_T + \frac{D}{2}
\label{eq:L_approx}
\end{equation}
where $D$ is the decrease rate of the variance of $\Lambda_T(\tau)$,
\[
D = \lim_{\tau \to \infty} \left[   
\langle \Lambda_T^2(\tau) \rangle - \lambda_T^2
 \right] \tau 
\]
From the data reported in Fig.~\ref{fig:lyap_t}, which refer to three different ensemble sizes, it is clear that the diffusion
rate vanishes for $\tau \to \infty$. According to Eq.~(\ref{eq:L_approx}) we can conclude that
$\mathcal{L}(1) = \mathcal{L}(0)$ and thereby that both vanish.
So this is a special case, where fluctuations do not induce multifractal fluctuations (at least at the perturbative level) and, more
important, the observation of a non-clustered distribution is consistent with the stability analysis.

\subsection{Entropy}

\begin{figure}[t]
  \centerline{\includegraphics[width=0.45\textwidth]{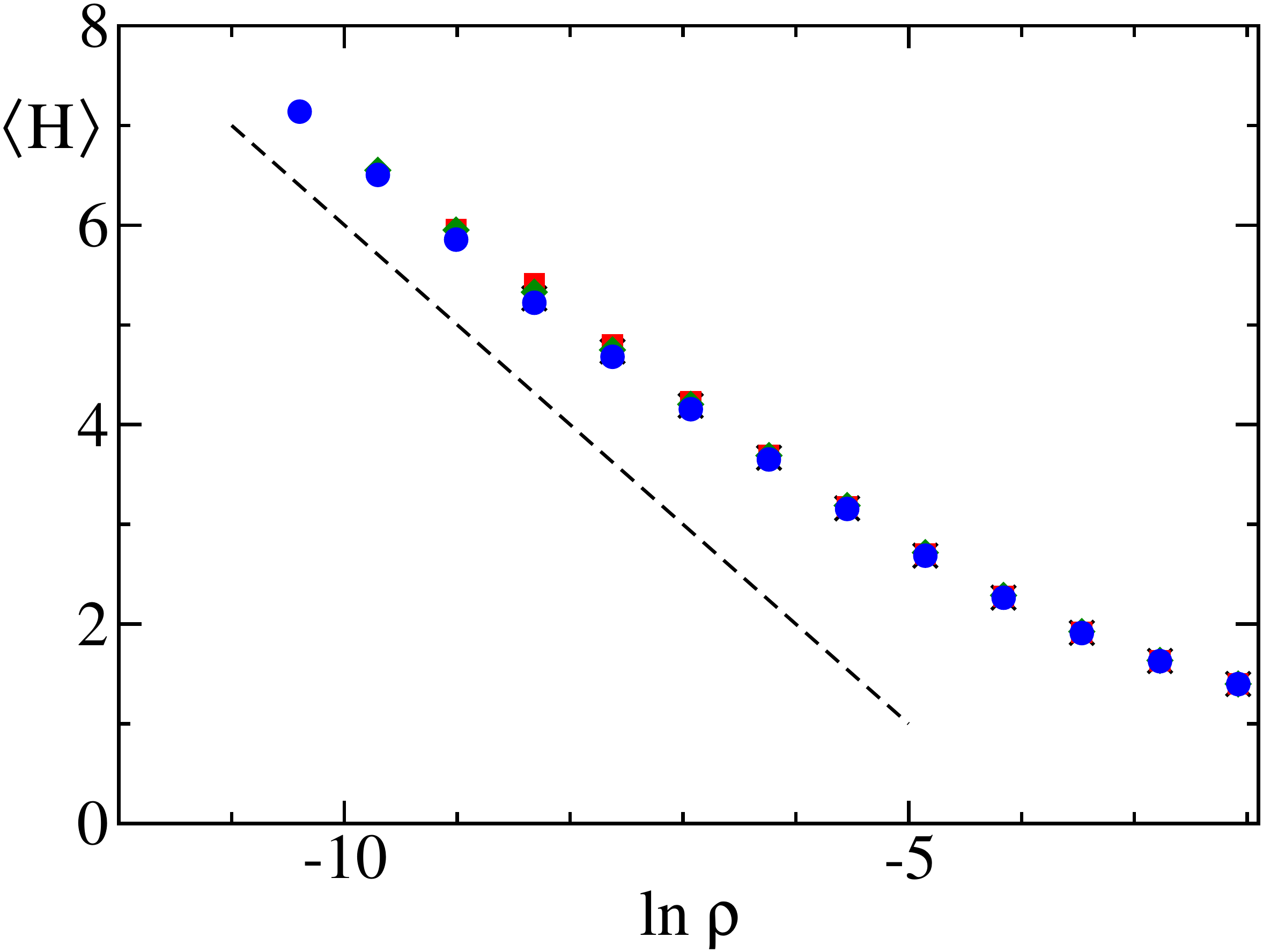}}
  \caption{Average entropy of the phase distribution versus angular resolution $\rho$ for four different ensemble sizes (crosses
correspond to $N=4096$, squares to $N=8192$, diamonds to $N=16384$, and circles to $N=32768$).
The straight line corresponds to a $-\!\ln \rho$ scaling.}
        \label{fig:entropy_scale}
\end{figure}

What about the structure of the distribution of phases? Is it either fractal or smooth? We have investigated its average
properties by computing the family of entropies introduced in section~\ref{methods:entropy}.
In Fig.~\ref{fig:entropy_scale}, we plot the average $\langle H\rangle$
versus $\ln \rho$ for four system sizes. The various  data sets start for different $\rho$ values, since
the smallest accessible scale being $\rho_m= 1/N$ depends on $N$.
The very good data collapse suggests that finite-size corrections are negligible over the resolutions accessible to the
numerical simulations. In other words,
we do not expect variations for the $\rho$ values larger than $\textrm{e}^{-10} \approx 4.5 \cdot 10^{-5}$.
The dependence of $H$ on $\rho$ gives information on the possibly fractal structure of the distribution: a scaling
behavior $H \propto -D_F \log \rho$ for $\rho \to 0$ would suggest a fractal dimension equal to $Di_F$.
The comparison with the dashed line, which corresponds to $\langle H \rangle = - \ln \rho + \textrm{const.}$,
 indicates
that the effective (resolution dependent) dimension $D_F(\rho)$ approaches 1 at small scales.
This result is consistent with the hypothesis that regime A is not a multi-cluster state.
However, the convergence towards a slope -1 is slow. The smoothness of the distribution can
be appreciated only for $\rho$ below $4.5\cdot 10^{-5}$. This explains the difficulties  encountered
in our simulations and in particular the difficulty one expects to encounter while integrating 
directly the equation for the probability density (a Liouville-like operator).

\section{Discussion and open problems}

Identifying the types  of collective behavior that can 
arise in mean-field models of identical oscillators 
is a goal of primary interest, even when focussing on the fairly restricted 
class of networks of identical phase oscillators such as in this paper.
In fact, this is a necessary step to eventually understand the true role played by ingredients
such as delay, heterogeneity, or an increased dimensionality of the single oscillators, encountered
in more realistic setups.

In this paper we have discussed a macroscopically chaotic dynamics, which arises in a model of phase-oscillators of
Kuramoto-Daido type with three harmonics.
It is a close analogon of the collective chaos analysed in Ref.~\cite{Clusella2019}, while studying a
mean-field model of Stuart-Landau oscillators. However, this is the first evidence of
one such a type of regime in oscillators characterized by a single variable: their phase.
The main difference between the two regimes is their transversal stability.
In the triharmonic model, the two generalized (transversal) Lyapunov exponents $\mathcal{L}(0)$ and $\mathcal{L}(1)$ 
vanish, suggesting that multifractal fluctuations are irrelevant, 
while $\mathcal{L}(1)=0 > \mathcal{L}(0)$ in Ref.~\cite{Clusella2019}.

Additionally, the chaotic regime emerging in the triharmonic model in finite systems is a transient, although it
lasts so long (at least $10^6$ times longer than the period
of the single oscillator) that it makes sense to perform an analysis as if it was strictly stationary.
Our studies, carried out for different numbers of oscillators (up to 32768), suggest that 
relevant observables such as the power spectrum or the distribution entropy are substantially independent of the system size
over the explored time scales ($10^6$ time units).

More tricky is the time-dependence over the same time span (see the evolution of $\exp(h)$), 
which challenges the accuracy of some results such as the value of the maximum Lyapunov exponent. 
The analysis of the phase-differences $\delta_i$ suggests, however, that the lack of stationarity affects mainly the
distribution of phases over small scales. More detailed simulations are required to draw firm conclusions.
In principle, this can be achieved by integrating the macroscopic evolution equation for the probability density.
However, our simulations of the single oscillators suggest that in order to fully resolve the quasi-cluster states,
it would be necessary to consider grids of at least $10^5$ points. 

The addition of a small amount of noise (transforming 
the evolution equation into a nonlinear Fokker-Planck equation)
can  regularize the density and soften the stringent constraints on the grid size. 
The chaotic regime discussed in this paper is, however, fragile and it would be perhaps preferable to explore a wider region of the parameter space to
identify a more robust instance of this collective chaos. 

\section*{Acknowledgements}

P.C. acknowledges financial support from the Spanish MINECO Project No.
FIS2016-76830-C2-1-P.

\section*{References}
\bibliographystyle{unsrt}
\bibliography{references}

\end{document}